\documentclass[10pt,conference]{IEEEtran}

\usepackage[T1]{fontenc}%
\usepackage{amsmath,amsfonts,amssymb,amsbsy,amstext,amsthm}
\usepackage{color}
\usepackage{tabularx}
\usepackage[ruled]{algorithm}
\usepackage{algorithmic}
\usepackage{cite}
\usepackage{bm}
\newcommand{\mat}[1]{\mbox{\boldmath{$#1$}}}

\usepackage{color}
\definecolor{orange}{rgb}{1,0.5,0}

\usepackage[cmintegrals]{newtxmath}

\usepackage[display-foreign=false]{acro}

\NewDocumentCommand{\acro}{m o m o}
{%
	\IfValueTF{#2}{%
		\IfValueTF{#4}{%
			\DeclareAcronym{#1}{short={#2},long={#3},#4}
		}{%
			\DeclareAcronym{#1}{short={#2},long={#3}}
		}
	}{%
		\IfValueTF{#4}{%
			\DeclareAcronym{#1}{short={#1},long={#3},#4}
		}{%
			\DeclareAcronym{#1}{short={#1},long={#3}}
		}
	}
}

\acro{AR}{autoregressive}
\acro{BF}{block fading}
\acro{ARMA}{autoregressive moving average}
\acro{PDPC}{pilot-and-data power control}
\acro{DSO}{Digital Switchover}
\acro{TV}{Television}
\acro{PLF}{predecessor leader following}
\acro{WHO}{world health organization}
\acro{PL}{platoon leader}
\acro{LTE-V2V}{long term evolution V2V}
\acro{PM}{platoon member}
\acro{CAM}{cooperative awareness message}
\acro{D2D}{device-to-device}
\acro{AWGN}{additive white Gaussian noise}
\acro{TVWS}{\acs*{TV} White Spaces}
\acro{RSRP}{Reference Signal Received Power}
\acro{RSRQ}{Reference Signal Received Quality}
\acro{RSSI}{Received Signal Strength Indicator}
\acro{SWOT}{Strengths, Weaknesses, Opportunities, and Threats}
\acro{AP}{access point}
\acro{MS}{mobile station}
\acro{ID}{Identity}
\acro{NCL}{Neighbor Cell List}
\acro{WLAN}{Wireless Local Area Network}
\acro{WiMAX}{Worldwide Interoperability for Microwave Access}
\acro{LTE}{long term evolution}
\acro{DMRS}{Demodulation Reference Signal}
\acro{UL}{uplink}
\acro{DL}{downlink}
\acro{PCI}{Physical-layer Cell \acs*{ID}}
\acro{TTI}{transmission time interval}
\acro{CSI}{channel state information}
\acro{ISM}{Industrial, Scientific and Medical}
\acro{3GPP}{3rd generation partnership project}
\acro{M2M}{Machine-to-Machine}
\acro{MTC}{Machine Type Communication}
\acro{eNB}{evolved node B}
\acro{PRB}{Physical Resource Block}
\acro{PF}{Proportional Fair}
\acro{RRA}{radio resource allocation}
\acro{MCS}{modulation and coding scheme}
\acro{SCM}{Spatial Channel Model}
\acro{SNR}{signal-to-noise ratio}
\acro{PDF}{probability density function}
\acro{GPS}{Global Positioning System}
\acro{P2P}{Peer to Peer}
\acro{MAC}{Medium Access Control}
\acro{OFDM}{Orthogonal Frequency Division Multiplexing}
\acro{CCL}{Common Cell List}
\acro{5G}{fifth generation}
\acro{FDD}{frequency division duplex}
\acro{TDD}{time division duplex}
\acro{EPA}{Equal Power Allocation}
\acro{MR}{Maximum Rate}
\acro{PhD}{Doctor of Philosophy}
\acro{3G}{Third Generation}
\acro{4G}{Fourth Generation}
\acro{ANACOM}{Autoridade Nacional de Comunicações}
\acro{AF}{array factor}
\acro{ARPU}{Average Return Per User}
\acro{AHP}{Analytic Hierarchy Process}
\acro{B3G}{Beyond \acs*{3G}}
\acro{BSc}{Bachelor of Science}
\acro{BV}{beamforming vector}
\acro{BWA}{Broadband Wireless Access}
\acro{CAPEX}{Capital Expenditure}
\acro{COGEU}{COgnitive radio systems for efficient sharing of TV white spaces in EUropean context}
\acro{CR}{Cognitive Radio}
\acro{CV}{combining vector}
\acro{DEnKF}{deterministic ensemble Kalman filter}
\acro{DFT}{discrete Fourier transform}
\acro{DSRC}{dedicated short-range communication}
\acro{DVB-H}{Handled Digital Video Broadcast}
\acro{DVB-T}{Terrestrial Digital Video Broadcast}
\acro{DTT}{Digital Terrestrial \acs*{TV}}
\acro{ETSI}{European Telecommunications Standards Institute}
\acro{FCC}{Federal Communications Commission}
\acro{FP7}{Seventh Framework Programme}
\acro{HM}{high mobility}
\acro{HSPA}{High Speed Packet Access}
\acro{IA}{initial access}
\acro{ICT}{Information and Communication Technologies}
\acro{IEEE}{Institute of Electrical and Electronics Engineers}
\acro{DySPAN-SC}{\acs*{IEEE} DySPAN Standards Committee}
\acro{ITU}{International Telecommunications Union}
\acro{LAN}{Local Area Network}
\acro{LS}{least squares}
\acro{MBMS}{Multimedia Broadcast Multicast Services}
\acro{MVNO}{Mobile Virtual Network Operator}
\acro{MT}{mobile terminal}
\acro{MSc}{Master of Science}
\acro{NPV}{Net Present Value}
\acro{NRA}{National Regulator Authority}
\acro{OFCOM}{Office of Communications}
\acro{OFDMA}{orthogonal frequency division multiple access}
\acro{OMP}{orthogonal matching pursuit}
\acro{OPEX}{Operational Expenditure}
\acro{PHY}{Physical}
\acro{PMSE}{Programme Making and Special Events}
\acro{PT}{Portugal}
\acro{RRS}{Reconfigurable Radio Systems}
\acro{QoE}{Quality of Experience}
\acro{SAE}{System Architecture Evolution}
\acro{SDR}{Software Defined Radio}
\acro{SLA}{Service Level Agreement}
\acro{SVD}{singular-value decomposition}
\acro{TETRA}{Trans European Trunked Radio Access}
\acro{ULA}{uniform linear array}
\acro{UHF}{Ultra High Frequency}
\acro{UK}{United Kingdom}
\acro{UMTS}{Universal Mobile Telecommunications System}
\acro{UPA}{uniform planar array}
\acro{UTRA}{\acs*{UMTS} Terrestrial Radio Access}
\acro{USP}{Unique Selling Point}
\acro{USA}{United States of America}
\acro{WG}{Working Group}
\acro{Wi-Fi}{Wireless Fidelity}
\acro{WRC}{World Radiocommunication Conference}
\acro{WSD}{White Space Device}
\acro{OM}[O\&M]{Operational \& Maintenance}
\acro{IRR}{Internal Rate of Return}
\acro{EKF}{extended Kalman filter}
\acro{EnKF}{ensemble Kalman filter}
\acro{EPC}{Evolved Packet Core}
\acro{E-UTRAN}{Evolved Universal Terrestrial Radio Access Network}
\acro{MME}{Mobility Management Entity}
\acro{PCRF}{Policy and Charging Rules Function}
\acro{S-GW}{Service Gateway}
\acro{P-GW}{Packet Gateway}
\acro{EIRP}{Equivalent Isotropically Radiated Power}
\acro{CBR}{Constant Bit Rate}
\acro{RRM}{Radio Resource Management}
\acro{HetNet}{Heterogeneous Network}
\acro{CEPT}{European Conference of Postal and Telecommunications Administrations}
\acro{ANATEL}{Agência Nacional de Telecomunicações}
\acro{STTD}{Space Time Transmit Diversity}[long-plural-form={Space Time Transmit Diversities}]
\acro{RAT}{Radio Access Technology}
\acro{RAN}{Radio Access Network}
\acro{SC}{Single Connection}
\acro{FS}{Fast Switching}
\acro{DC}{Dual Connectivity}
\acro{C-RAN}{Cloud \acs*{RAN}}
\acro{NR}{new radio}
\acro{mmW}{millimeter wave}
\acro{KF}{Kalman filter}
\acro{KPI}{Key Performance Indicator}
\acro{BB}{branch and bound}
\acro{BER}{bit error rate}
\acro{BS}{base station}
\acro{CDF}{cumulative distribution function}
\acro{CFN}{channel Frobenius norm}
\acro{DCP}{difference of convex functions program}
\acro{GTEL}{Wireless Telecommunications Research Group}
\acro{GUB}{global upper bound}
\acro{GLB}{global lower bound}
\acro{HMSINR}{highest minimum \ac{SINR}}
\acro{HPBW}{half power beamwidth}
\acro{HWCR}{highest \ac{WCR}}
\acro{HWSR}{highest weighted sum rate}
\acro{IBC}{interference broadcast channel}
\acro{IC}{interference channel}
\acro{KKT}{Karush-Kuhn-Tucker}
\acro{LB}{lower bound}
\acro{LiDAR}{light detection and ranging}
\acro{LOS}{line of sight}
\acro{LP}{linear programming}
\acro{LP-MMSE}{local partial MMSE}
\acro{MAP}{a maximum a posteriori probability}
\acro{MILP}{mixed integer linear programming}
\acro{MIMO}{multiple input multiple output}
\acro{MINLP}{mixed integer nonlinear programming}
\acro{MIP}{mixed integer programming}
\acro{MISO}{multiple input single output}
\acro{ML}{machine learning}
\acro{MMSE}{minimum mean squared error}
\acro{MRC}{maximum ratio combining}
\acro{MRT}{maximum ratio transmission}
\acro{MSE}{mean square error}
\acro{MU-MIMO}{multi-user multiple input multiple output}
\acro{MU}{multi-user}
\acro{NLP}{nonlinear programming}
\acro{NLOS}{non-line of sight}
\acro{NMSE}{normalized mean square error}
\acro{NP}{non-polynomial time}
\acro{QoS}{quality of service}
\acro{SCA}{successive convex approximation}
\acro{SINR}{signal-to-interference-plus-noise ratio}
\acro{SIR}{signal-to-interference ratio}
\acro{SISO}{single input single output}
\acro{SIMO}{single input multiple output}
\acro{SOCP}{second-order cone programming}
\acro{SV}{steering vector}
\acro{TU}{typical urban}
\acro{UE}{user equipment}
\acro{UKF}{unscented Kalman filter}
\acro{UB}{upper bound}
\acro{WINNER}{Wireless World Initiative New Radio}
\acro{WMMSE}{weighted minimum mean squared error}
\acro{WCR}{weighted common rate}
\acro{V2V}{vehicle-to-vehicle}
\acro{V2I}{vehicle-to-infrastructure}
\acro{V2P}{vehicle-to-pedestrian}
\acro{V2X}{vehicle-to-everything}
\acro{RF}{radio frequency}
\acro{RL}{reinforcement learning}
\acro{RRV}{rate-relaxation variable}
\acro{DQL}{deep Q-learning}
\acro{DQN}{deep Q-network}
\acro{ADMM}{alternating direction method of multipliers}
\acro{SDP}{semidefinite programming}
\acro{PARAFAC}{parallel factor analysis}
\acro{V2N}{vehicle-to-network}
\DeclareAcronym{AoD}{
    short = AoD,
    long = angle of departure,
    long-plural-form = angles of departure
}
\DeclareAcronym{AoA}{
    short = AoA,
    long = angle of arrival,
    long-plural-form = angles of arrival
}
\acro{BALS}{bilinear alternating least-squares}
\acro{TALS}{trilinear alternating least-squares}
\acro{RMSE}{root mean square error}
\acro{ALS}{alternating least-squares}
\acro{5G-PPP}{5G infrastructure public private partnership}
\acro{SLNR-MAX}{maximum signal-to-leakage-plus-noise ratio}
\acro{SLNR}{signal-to-leakage-plus-noise ratio}
\acro{SUS}{signal-to-leakage-plus-noise ratio based user selection}
\acro{ReLU}{rectifier linear unit}
\acro{ZF}{zero-forcing}

\acro{DB}{database}
\acro{EE}{energy efficiency}
\acro{KNN}{K-nearest neighbors}
\acro{PCA}{principal component analysis}
\acro{CIR}{channel impulse response}
\acro{CPU}{central processing unit}
\acro{mMIMO}{massive multiple input multiple output}
\acro{i.i.d.}{independent and identically distributed}
\acro{GP}{geometric programming}
\acro{SE}{spectral efficiency}
\acro{PDPR}{pilot-data power ratio} 

\DeclareMathAlphabet{\mathppl}{T1}{ppl}{m}{it}
\DeclareMathAlphabet{\mathphv}{T1}{phv}{m}{it}
\DeclareMathAlphabet{\mathpzc}{T1}{pzc}{m}{it}

\usepackage{graphicx}
\usepackage{caption}
\newtheorem{defi}{Definition}

\newtheorem{cor}{Corollary}
\newtheorem{lem}{Lemma}

\IEEEoverridecommandlockouts\IEEEpubid{\makebox[\columnwidth]{ 978-1-6654-5975-4/22~\copyright~2022 IEEE \hfill} \hspace{\columnsep}\makebox[\columnwidth]{ }}
\begin{document}

\title{A Game-Theoretic Solution for Power Management in the Uplink of Cell-Free Massive MIMO Systems}
\title{A Distributed Game-Theoretic Solution for Power Management for the Uplink of Cell-Free Systems}
\title{A Distributed Game-Theoretic Solution for Power Management in the Uplink of Cell-Free Systems}

\author{ Juno V. Saraiva, Roberto P. Antonioli, G\'abor Fodor, Iran M. Braga Jr., Walter C. Freitas Jr., and Yuri C. B. Silva%
}

\author{
	Juno V. Saraiva$^{\star}$,
	Roberto P. Antonioli$^{\star,*}$,
	G\'{a}bor Fodor$^{\dag,\ddag}$,
	Walter C. Freitas Jr.$^\star$,
	Yuri C. B. Silva$^\star$\\
	\small Wireless Telecom Research Group (GTEL)$^\star$, Federal University of Cear\'a, Fortaleza, Brazil.\\
	\small Instituto Atlântico$^*$, Fortaleza, Brazil.
	\small Ericsson Research$^\dag$ and Royal Institute of Technology$^\ddag$, Stockholm, Sweden. \\
	\small \{juno, antonioli, walter, yuri\}@gtel.ufc.br, roberto\_antonioli@atlantico.com.br, gabor.fodor@ericsson.com, gaborf@kth.se
}

\maketitle

\acresetall

\IEEEpeerreviewmaketitle

\begin{abstract}
This paper investigates cell-free massive multiple input multiple output systems with a particular focus on uplink power allocation. In these systems, uplink power control is highly non-trivial, since a single user terminal is associated with multiple intended receiving base stations. In addition, in cell-free systems, distributed power control schemes that address the inherent spectral and energy efficiency targets are desirable. By utilizing tools from game theory, we formulate our proposal as a non- cooperative game, and using the best-response dynamics, we obtain a distributed power control mechanism. To ensure that this power control game converges to a Nash equilibrium, we apply the theory of potential games. Differently from existing game- based schemes, interestingly, our proposed potential function has a scalar parameter that controls the power usage of the users. Numerical results confirm that the proposed approach improves the use of the energy stored in the battery of user terminals and balances between spectral and energy efficiency.
\end{abstract}

\begin{IEEEkeywords}
	Cell-free systems, game theory, Nash equilib- rium, potential game, power control, radio resource allocation.
\end{IEEEkeywords}

\section{Introduction}

Radio resource allocation is a major issue in the design of modern mobile networks. In interference-limited systems, for example, power control plays an indispensable role in managing interference, ensuring proper signal strength at the intended receivers and saving energy.
Recognizing the scarcity of the energy resource and the growing worldwide energy concern, efficient power control solutions have definitely become a key requirement for the continued success of wireless systems~\cite{Miao2012,Liu2014, He_2017}. 
Especially related to the uplink and given the ever-increasing growth in mobile subscriptions, an efficient power allocation strategy is important to reduce energy demands and battery consumption.
By mitigating interference levels, power control has also the advantage of providing a more uniform throughput among users. 
Furthermore, an optimized energy consumption contributes to reducing environmental impacts, e.g., heat dissipation and electronic pollution~\cite{Miao2013}.

However, the power management in cellular networks is a fairly complex problem. 
In general, efficiently controlling power usage with multiple interfering users may lead to non-polynomial time (NP)-hard problems, and in these cases obtaining optimal solutions is extremely challenging. 
Normally, within the power allocation framework, the main difficulties in finding alternative solutions are the performance coupling among the users as well as their inherently selfish behaviors. 
Consequently, a good solution needs to deal with the interactions among several independent users with contrasting interests. 
In this context, game theory provides a natural framework for developing mechanisms when many individuals with conflicting interests interact. 
Therefore, it is a promising approach to study interactions among contending users in order to seek feasible and practically viable solutions~\cite{mackenzie2006game,Osborne2004}. 
Indeed, there has been growing interest in adopting game-theoretic methods to propose alternative solutions in mobile communications, see, e.g.,~\cite{Buzzi2012, Xie2014, Zhao_2018, Zhao2019,Juno_2022,Myung2022}. 

In~\cite{Buzzi2012} and~\cite{Xie2014}, the authors focused on an uplink power control game-based solutions for \ac{OFDMA} systems and cognitive radio networks, respectively. 
To address the problem of minimizing the sum of the \acp{MSE}, power control schemes for the uplink of massive \ac{MU-MIMO} systems were proposed in~\cite{Zhao_2018, Zhao2019, Juno_2022}. 
More specifically, considering block fading channels,~\cite{Zhao_2018} and~\cite{Zhao2019} relied on game theory to optimize the pilot-to-data power ratio assuming single and multi-cell cases, respectively. 
Likewise, a game-based approach to controlling the pilot and data power levels was presented in~\cite{Juno_2022} while considering more realistic auto-regressive channels. 
On the other hand, in more recent network architectures such as cell-free, game-theoretic approaches to radio resource allocation have {not been widely explored} in the literature, mainly related to the uplink. Recently, in~\cite{Myung2022}, a power control game was proposed for cell-free massive \ac{MIMO} systems, but the authors focused on the downlink.
 
Inspired by the above discussion, this paper considers the problem of power control for the uplink of cell-free massive \ac{MIMO} systems. 
Due to the inherent competitive nature of the multi-user and user-centric environment, we use a game-theoretic framework and model the problem as a strategic non-cooperative game, which can often provide feasible and convenient alternatives for a distributed implementation. 
Specifically, we use novel payoff functions based on an adapted \ac{SIR} expression. 
More importantly, different from existing works, our solution is designed as a parameterized potential game for which the existence and uniqueness of a Nash equilibrium is ensured.
Thereby, we show that the proposed power control achieves efficient solutions with respect to different network objectives such as sum-rate maximization, max-min fairness or power consumption minimization.  

\section{Network Model}
\label{Sec:Network_Model}

We consider a cell-free massive \ac{MIMO} system consisting of $K$ single-antenna \acp{UE} and $L$ \acp{AP} equipped with $N$ antennas grouped in the sets $\mathcal{K}$ and $\mathcal{L}$, respectively. 
The \acp{AP} and \acp{UE} are deployed randomly in a wide area without boundaries. 
A central processing unit (CPU) connects with the \acp{AP} via a backhaul network.

Particularly, we analyze a cell-free massive \ac{MIMO} system operating in \ac{TDD} mode with a pilot phase for channel estimation and a data transmission phase. 
Each coherence block is divided into $\tau_{\text{p}}$ channel uses for uplink pilots, $\tau_{\text{u}}$ for uplink data and $\tau_{\text{d}}$ for downlink data such that $\tau_{\mathrm{c}} = \tau_{\mathrm{p}} + \tau_{\mathrm{u}} + \tau_{\mathrm{d}}$. 
The channel between \ac{AP} $l$ and \ac{UE} $k$ is denoted as $\mathbf{h}_{kl}\in \mathbb{C}^{N}$ and $\mathbf{h}_k = \left[\mathbf{h}^{\mathrm{T}}_{k1}, \dots, \mathbf{h}^{\mathrm{T}}_{kL}\right]^{\mathrm{T}} \in \mathbb{C}^{NL}$ is the collective channel from all \acp{AP}. 
In each coherence block, an independent realization from a correlated Rayleigh fading distribution is drawn as $\mathbf{h}_{kl} \sim \mathcal{N}_{\mathbb{C}}(\mathbf{0}, \mathbf{R}_{kl})$, where $\mathbf{R}_{kl}$ is the spatial correlation matrix describing the spatial property of the channel and $\beta_{kl} =\mathrm{tr}(\mathbf{R}_{kl})/N$ is the large-scale fading coefficient that describes pathloss and shadowing \cite{Emil_Luca_2020, Chen_2021}. 
The Gaussian distribution models the small-scale fading whereas the positive semi-definite correlation matrix $\mathbf{R}_{kl}$ describes the large-scale fading, including shadowing, pathloss, spatial channel correlation and antenna gains. 
Given that the \acp{AP} are spatially distributed in the system, the channel vectors of different \acp{AP} are independently distributed, i.e., $\mathbb{E}\left\{\mathbf{h}_{kl^{'}} \left(\mathbf{h}_{kl}\right)^{\mathrm{H}} \right\} = \mathbf{0}$ when $l^{'} \neq l$. 
The collective channel is distributed as $\mathbf{h}_{k} \sim \mathcal{N}_{\mathbb{C}}(\mathbf{0}, \mathbf{R}_{k})$, where $\mathbf{R}_k = \mathrm{diag}\left(\mathbf{R}_{k1}, \dots, \mathbf{R}_{kL} \right) \in \mathbb{C}^{NL\times NL}$ is the block-diagonal spatial correlation matrix~\cite{Emil_Luca_2020}. 

We define a set of block-diagonal matrices $\mathbf{D}_k = \mathrm{diag}\left(\mathbf{D}_{k1}, \dots, \mathbf{D}_{kL} \right) \in \mathbb{C}^{NL\times NL}, \, k \in \mathcal{K}$, where $\mathbf{D}_{il}\in \mathbb{C}^{N\times N}, \, i \in \mathcal{K} \text{ and } l \in \mathcal{L}$ is the set of diagonal matrices, determining which \ac{AP} antennas may transmit to which \acp{UE}.
More specifically, the $n$-th diagonal entry of $\mathbf{D}_{il}$ is 1 if the $n$-th antenna of \ac{AP} $l$ is allowed to transmit and to decode signals from \ac{UE} $k$, and 0 otherwise. 
Based on the definition of the set of matrices $\mathbf{D}_{il}$,  we define a matrix $\mathbf{A}\in \mathbb{R}^{K\times L}$ specifying the \ac{AP} selection, where $A_{k,l}=1$ if \ac{AP} $l$ is allowed to transmit and to decode signals from \ac{UE} $k$, i.e., if $\mathrm{tr}\left(\mathbf{D}_{kl} \right) > 0$, and $0$ otherwise. For the conciseness of mathematical descriptions, we denote by $\mathcal{M}_k = \{l \,\big\rvert \, A_{k,l}=1, l\in \mathcal{L}\}$ the subset of \acp{AP} serving \ac{UE} $k$. Meanwhile, $\mathcal{D}_l =  \{k \,\big\rvert \, A_{k,l}=1, k\in \mathcal{K}\}$ is the subset of \acp{UE} served by \ac{AP} $l$.

\subsection{Uplink pilot-based channel estimation}

We consider that there are $\tau_{\text{p}}$ mutually orthogonal $\tau_{\text{p}}$-length pilots, with $\tau_{\text{p}}$ being a constant independent of $K$. Let $\mathcal{S}_t \subset \mathcal{K}$ be the subset of \acp{UE} assigned to pilot $t$. When the \acp{UE} in $\mathcal{S}_t$ transmit, the received signal $	\mathbf{y}^{\mathrm{pilot}}_{tl} \in \mathbb{C}^{N}$ at \ac{AP} $l$ is   
\begin{align}
	\label{signal_uplink}
	\mathbf{y}^{\mathrm{pilot}}_{tl} = \sum_{i \in \mathcal{S}_t}\sqrt{\tau_{\mathrm{p}}\rho_i}\mathbf{h}_{il} + \mathbf{n}_{tl},	
\end{align}
where $\rho_i$ is the transmit power of \ac{UE} $i$, $\tau_{\mathrm{p}}$  is the processing gain, and $\mathbf{n}_{tl} \sim \mathcal{N}_{\mathbb{C}}(\mathbf{0}, \sigma^{2}\mathbf{I}_N)$ is the thermal noise.  Note that, since we assume a massive access scenario with a large number of \acp{UE}, i.e., $K>\tau_{\text{p}}$, several \acp{UE} share the same pilot as shown in~\eqref{signal_uplink}, leading to pilot contamination.

For estimating the channels, the classic \ac{MMSE} criterion has been recurrently employed in the literature. 
The \ac{MMSE} estimate of $\mathbf{h}_{kl}$ for \ac{UE} $k \in \mathcal{S}_t$ is $\hat{\mathbf{h}}_{kl} = \sqrt{\tau_{\mathrm{p}}\rho_k}\mathbf{R}_{kl}\mat{\Psi}_{tl}^{-1}\mathbf{y}^{\mathrm{pilot}}_{tl}$, where $\mat{\Psi}_{tl} =  \sum_{i \in \mathcal{S}_t} \tau_{\mathrm{p}}\rho_i \mathbf{R}_{il} + \sigma^{2}\mathbf{I}_{N}$
is the correlation matrix of~\eqref{signal_uplink}. 
The estimated channel $\hat{\mathbf{h}}_{kl}$ and estimation error $\tilde{\mathbf{h}}_{kl}  = \mathbf{h}_{kl} - \hat{\mathbf{h}}_{kl}$ are independent vectors distributed as $\hat{\mathbf{h}}_{kl} \sim \mathcal{N}_{\mathbb{C}}(\mathbf{0}, \mathbf{B}_{kl})$ and $\tilde{\mathbf{h}}_{kl} \sim \mathcal{N}_{\mathbb{C}}(\mathbf{0}, \mathbf{C}_{kl})$, where 
$\mathbf{B}_{kl} = \mathbb{E}\left\{ 	\hat{\mathbf{h}}_{kl} 	\hat{\mathbf{h}}^{\mathrm{H}}_{kl}   \right\} = \tau_{\mathrm{p}}\rho_k\mathbf{R}_{kl}\mat{\Psi}_{tl}^{-1}\mathbf{R}_{kl}$ and $\mathbf{C}_{kl} = \mathbb{E}\left\{ 	\tilde{\mathbf{h}}_{kl} 	\tilde{\mathbf{h}}^{\mathrm{H}}_{kl}	\right\} = \mathbf{R}_{kl} - \mathbf{B}_{kl}.$

\subsection{Uplink data transmission}
During the uplink data transmission, \ac{AP} $l$ receives the signal $\mathbf{y}_{l}\in \mathbb{C}^{N}$ from all \acp{UE}, as
\begin{align}
\mathbf{y}_{l} = \sum_{k \in \mathcal{K}}\mathbf{h}_{kl}s_{k} + \mathbf{n}_{l},
\end{align}
where $s_{k} \in \mathbb{C}$ is the signal transmitted from \ac{UE} $k$ with power $\rho_k$ and $\mathbf{n}_{l} \sim \mathcal{NC}(\mathbf{0}, \sigma^{2}\mathbf{I}_N)$. However, since only a subset of \acp{AP} take part in the signal detection,  the estimate of $s_k$ is:
\begin{align}
	\hat{s}_k &= \sum_{l \in \mathcal{L}}\mathbf{v}^{\mathrm{H}}_{kl}\mathbf{D}_{kl}\mathbf{y}_{l}	\nonumber\\
	& =   \mathbf{v}^{\mathrm{H}}_{k}\mathbf{D}_{k}\mathbf{h}_{k}s_k + \sum_{i \in \mathcal{K}\backslash \{k\}}\mathbf{v}^{\mathrm{H}}_{k}\mathbf{D}_{k}\mathbf{h}_{i}s_i + \mathbf{v}^{\mathrm{H}}_{k}\mathbf{D}_{k}\mathbf{n},
\end{align}
 where $\mathbf{v}_{kl}\in \mathbb{C}^{N}$ is a receive combining vector of \ac{AP} $l$ for \ac{UE} $k$, $\mathbf{v}_k =   \left[\mathbf{v}^{\mathrm{T}}_{k1}, \dots, \mathbf{v}^{\mathrm{T}}_{kL}\right]^{\mathrm{T}} \in \mathbb{C}^{NL}$ denotes the collective of these combining vectors and $\mathbf{n} =   \left[\mathbf{n}^{\mathrm{T}}_{1}, \dots, \mathbf{n}^{\mathrm{T}}_{L}\right]^{\mathrm{T}} \in \mathbb{C}^{NL}$ collects all the noise vectors.
 
Preferably, for large-scale networks, it is more interesting to direct the main computational tasks to the \acp{AP} in a distributed way and, thus, avoid overloading the CPU. Therefore, instead of sending $\{	\mathbf{y}^{\mathrm{pilot}}_{tl}\}_{\forall t}$ and $\mathbf{y}_{l}$ to the CPU, each \ac{AP} $l$ locally selects the combining vector $\mathbf{v}_{kl}$ and then it preprocesses its signal by computing local estimates of the data as $\hat{s}_{kl} = \mathbf{v}^{\mathrm{H}}_{kl}\mathbf{D}_{kl}\mathbf{y}_{l}$. Next, the local estimates of all \acp{AP} that serve \ac{UE} $k$ are sent to the CPU for final estimate of $s_k$, which is given by $\hat{s}_{k} = \sum_{l \in \mathcal{L}}\hat{s}_{kl}$. We utilize the \textit{use-and-then-forget} bound to obtain the achievable \ac{SE}.
\begin{lem}~\textnormal{\cite{Emil_Luca_2020, Chen_2021}}.
	\label{lem:SINR}
	An achievable uplink \ac{SE} for \ac{UE} $k$ is 
	\begin{align}
		\mathrm{SE}_k = \frac{\tau_{\mathrm{u}}}{\tau_{\mathrm{c}}}\log_{2}(1+\mathrm{SINR}_k), 
	\end{align}
	where
	\begin{align}
		\label{eq:SINR}
		&\mathrm{SINR}_k = \nonumber \\ &\frac{\rho_k\big\vert \mathbb{E}\{ \mathbf{v}_k^{\mathrm{H}} \mathbf{D}_k \mathbf{h}_k \} \big\vert^{2}}{\sum\limits_{i \in \mathcal{K}}\rho_i \mathbb{E}\left\{\big\vert \mathbf{v}_k^{\mathrm{H}} \mathbf{D}_k \mathbf{h}_i\big\vert^{2} \right\} - \rho_k\Big\vert\mathbb{E}\{ \mathbf{v}_k^{\mathrm{H}} \mathbf{D}_k \mathbf{h}_k \} \Big\vert^{2} + \sigma^{2}\mathbb{E}\left\{\big\vert\big\vert\mathbf{D}_k \mathbf{v}_k\big\vert\big\vert^{2}\right\}}.
	\end{align}
\end{lem}

In general, any combining vector that depends on the local channel estimates and statistics can be used in the \ac{SINR} expression~\eqref{eq:SINR}, but the expectations in it cannot be computed in closed form for any set of values $\{\mathbf{v}_{kl}\}_{\forall k,l}$. With simpler combining vector structures, such as maximum ratio combining (MRC), i.e., when $\mathbf{v}_{kl}=	\hat{\mathbf{h}}_{kl}$, it is possible to obtain closed form expressions. 
Nevertheless, the performance of MRC is quite limited, and significant performance gains can be obtained when using combining vectors also with distributed structures but based on the MMSE criterion, such as local partial MMSE (LP-MMSE) combining~\cite{Emil_Luca_2020, Chen_2021}, whose combining vector $\mathbf{v}_{kl}^{\mathrm{LP-MMSE}}$ is:
\begin{align}
	\mathbf{v}_{kl}^{\mathrm{LP-MMSE}} = \rho_k\left(\sum_{i \in \mathcal{D}_l} \rho_i \left(\hat{\mathbf{h}}_{il} \hat{\mathbf{h}}^{\mathrm{H}}_{il} + \mathbf{C}_{il}\right) + \sigma^{2}\mathbf{I}_{N}\right)^{-1} \hat{\mathbf{h}}_{kl}. 
\end{align}

However, when using $\{\mathbf{v}_{kl}^{\mathrm{LP-MMSE}}\}_{\forall k,l}$ the \ac{SINR} expression in~\eqref{eq:SINR} can only be computed via Monte Carlo simulations~\cite{Emil_Luca_2020}.

\section{Resource Allocation}
\label{SEC:power_allocation}

In this section, we describe the radio resource management employed in the network that consists of two parts. In the first part, we simply adopt the algorithm for joint initial access, pilot assignment, and cluster formation proposed in~\cite[See Section V-A]{Emil_Luca_2020}. 
Then, in the second part, differently from~\cite{Emil_Luca_2020}, we pay special attention to power control and propose a game-theoretic model of the interactions among users assuming a distributed management framework, which is presented in details in the following subsections.
\subsection{Game Theoretic Approach}
\label{SUB_SEC:game_theoretic_approach}

In the context of game theory, the players are considered as entities with the ability of observation and reaction. 
For the proposed game model, in order to mitigate potential interference levels in the uplink and obtain a suitable data power profile, the players are the users themselves. We define the proposed non-cooperative game $\mathcal{G}$ as $\mathcal{G}=\left\{\mathcal{K}, \{\mathcal{P}_{k}\}_{\forall k}, \{\mu_{k}(\rho_{k},\mat{\rho}_{(-k)})\}_{\forall k}\right\},$
%
%
where $\mathcal{K}$ is the set of \acp{UE}, i.e., a finite set of players. 
For a given \ac{UE} $k$,  $\mathcal{P}_{k}= \left[ \rho_{\text{min}}, \rho_{\text{max}}\right]$ is a finite set of available strategies or actions, where $\rho_{\text{min}}>0$ and $\rho_{\text{max}}\leq P_{\text{max}}$ with $P_{\text{max}}$ being the maximum uplink data power. 
In the context of the proposed game, the data power value $\rho_k \in \mathcal{P}_{k}$ denotes the strategy chosen by \ac{UE} $k$ and $\mat{\rho}_{(-k)}$ denotes the strategies of all the \acp{UE} other than \ac{UE}~$k$.  
Therefore, $\mat{\rho} = (\rho_k, \mat{\rho}_{(-k)}) = \left[\rho_{1}, \cdots, \rho_{k}, \cdots, \rho_{K} \right]^{\mathrm{T}}\in \mathbb{R}^{K}$ represents the profile of data powers of all \acp{UE}, i.e., a power allocation strategy for the system. 
Moreover, $\mu_{k}(\rho_k, \mat{\rho}_{(-k)})$: $\mat{\Upsilon} \rightarrow \mathbb{R}$, is a real-valued utility/payoff function where, $\mat{\Upsilon} = \mathcal{P}_{1}\times \mathcal{P}_{2}\times \cdots \times \mathcal{P}_{K}$ is the strategy space. 
Note that, for every chosen strategy by \ac{UE} $k$, the power profile $(\rho_{k},\mat{\rho}_{(-k)})$ is associated with a payoff, i.e., $\mu_{k}(\rho_{k},\mat{\rho}_{(-k)})$. 
Thus, the payoff function quantifies the preferences of each \ac{UE} to a given action, provided the knowledge of others' actions. 

Typically, a non-cooperative game is a procedure where each player will selfishly choose an action that improves its own utility function given the current strategies of the other players. Then, a key issue when designing a game is the choice of the payoff function.  Specifically, for game $\mathcal{G}$ the independent actions of the \acp{UE} to set their power values should not only provide satisfactory individual solutions but should also mitigate potential interference levels in the network. Thereby,  {we design a payoff function that enables \acp{UE} to have lower data power levels while causing less interference in other \acp{UE}, given by:} 
\begin{align}
	\label{payoff_function}
	\mu_k(\alpha, \rho_k, \mat{\rho}_{(-k)}) = \gamma_k(\alpha, \rho_k, \mat{\rho}_{(-k)})  + 	\lambda_k(\alpha, \rho_k, \mat{\rho}_{(-k)}),
\end{align}
where
\begin{subequations}
	\label{gamma_lambda}
	\begin{align}
		&\gamma_k(\alpha, \rho_k, \mat{\rho}_{(-k)}) = \frac{\sum\limits_{\forall i \neq k}\rho_i\Big(\sum\limits_{l\in \mathcal{M}_i}\beta_{i,l}\Big)^\alpha}{\rho_k\Big(\sum\limits_{ l\in \mathcal{M}_k}\beta_{k,l}\Big)^\alpha}, \\
		&\lambda_k(\alpha,\rho_k, \mat{\rho}_{(-k)}) =  \rho_k\Big(\sum\limits_{ l\in \mathcal{M}_k}\beta_{k,l}\Big)^\alpha\sum_{\forall i \neq k}\frac{1}{\rho_i\Big(\sum\limits_{l\in \mathcal{M}_i}\beta_{i,l}\Big)^\alpha},
	\end{align}
\end{subequations}
and $\alpha$ is an input parameter of game $\mathcal{G}$. 

Particularly, the term $\gamma_k(\alpha, \mat{\rho})$ is based on the reciprocal of the SIR expression shown in~\cite[Section V, Equation (51)]{Chen_2021}. 
Initially, assuming the particular case where $\{\lambda_k(\alpha,\rho_k, \mat{\rho}_{(-k)})\}_{\forall k}=0$ and given $\alpha$ and $\mat{\rho}_{(-k)}$ fixed, it is easy to see that the best strategy or action for each \ac{UE} exists and it would be to minimize the payoff function in~\eqref{payoff_function} by choosing the highest possible value for the data power. 
However, we add the term $\lambda_k(\alpha, \mat{\rho})$ to the payoff function $\mu_k(\alpha, \mat{\rho})$ in order to make the decision of the \acp{UE} non-trivial and especially less selfish. 
In a general case, i.e., even if $\{\lambda_k(\alpha,\rho_k, \mat{\rho}_{(-k)})\}_{\forall k}\neq0$, it is possible to obtain a single value that minimizes~\eqref{payoff_function} {as it is a convex function in $\mathcal{P}_{k},\forall  k \in \mathcal{K}$}. {By solving $\partial \mu_{{k}}(\alpha, \rho_{{k}}, \mat{\rho}_{(-{k})})/ \partial  \rho_{{k}} = 0$, we can find the unique minimizer, $\rho^{*}_k$, of~\eqref{payoff_function}, as follows:} 
{ 
\begin{align}
	\label{EQ:best_power}
	\rho^{*}_k = \sqrt{\left({\sum\limits_{\forall i \neq {k}}\rho_i\Big(\sum\limits_{l\in \mathcal{M}_i}\beta_{i,l}\Big)^\alpha}\right)\left({\sum\limits_{\forall i \neq k}\frac{\Big(\sum\limits_{ l\in \mathcal{M}_k}\beta_{k,l}\Big)^{2\alpha}}{\rho_i\Big(\sum\limits_{l\in \mathcal{M}_i}\beta_{i,l}\Big)^\alpha}}\right)^{-1}}.
\end{align}
}

From the point of view of the \acp{UE}, the term $\lambda_k(\alpha, \mat{\rho})$ in~\eqref{payoff_function} represents a punishment for the \ac{UE} who decides to excessively increase the value of the chosen data power. As a result, $\lambda_k(\alpha, \mat{\rho})$ can reduce interference levels in the system and avoid a greedy power allocation strategy, i.e., $\rho_k=P_{\text{max}},\, \forall k \in \mathcal{K}$.

In game theoretic approaches, we are interested in finding a Nash equilibrium as a solution. 
A Nash equilibrium  is a strategy profile that satisfies the condition that no player can unilaterally improve its own payoff as shown in Definition~\ref{definition_1}:
\begin{defi}
	\label{definition_1}
	An $\epsilon$-Nash equilibrium of parameterized game $\mathcal{G}(\alpha)$ is achieved when the payoff function $\mu_{k}\left(\alpha, \mat{\rho}\right)$ is such that for all \ac{UE} $k$:
	\begin{align}
		\mu_{k}\left(\alpha, \mat{\rho}\right) \leq \mu_{k}\big(\alpha, \rho_{k},\mat{\rho}_{-(k)}\big) + \epsilon,\,\,  \rho_{k} \in \mathcal{P}_k.
	\end{align}
\end{defi}

However, games may have a large number of Nash equilibrium points or may not have any. 
Generally, finding or even characterizing the set of these equilibrium points in terms of existence or uniqueness is a difficult task. 
Fortunately, there is a particular case of non-cooperative games called \textit{potential games} for which the existence and uniqueness of a Nash equilibrium is ensured~{\cite{Scutari2006}}. 
Basically, in a potential game the incentive of all players to change their actions can be expressed by a global payoff function called \textit{potential function}. 
Mathematically, the proposed parameterized game $\mathcal{G}(\alpha)$ is a potential game if it complies with Definition~\ref{definition_2}~\cite{mackenzie2006game}. 
\begin{defi}
	\label{definition_2}
	If the proposed parameterized game $\mathcal{G}(\alpha)$  is a potential game, then there exists a function $u :  \mat{\Upsilon} \rightarrow \mathbb{R}$ such that $\forall k \in \mathcal{K}$ and $\forall \rho_{k}, \rho'_k \in \mathcal{P}_k$:
	\begin{align}
		&u(\alpha, \rho'_k, \mat{\rho}_{(-k)}) - u(\alpha, \rho_k, \mat{\rho}_{(-i)}) =\nonumber \\&\quad\quad\quad\quad\quad\quad\quad \mu_{k}(\alpha, \rho'_k, \mat{\rho}_{(-i)}) - \mu_{k}(\alpha, \rho_k, \mat{\rho}_{(-k)}).
	\end{align}
	In this case, the function $u(\cdot)$ is called an exact potential function for the parameterized game $\mathcal{G}(\alpha)$~{\textnormal{\cite{mackenzie2006game,Osborne2004}}}.
\end{defi}

Note that based on Definition~\ref{definition_2}, $\mathcal{G}(\alpha)$ is a potential game if it is possible to define a potential function, i.e., an \ac{UE}-independent function that measures the same amount of change or marginal payoff for any unilaterally deviating \ac{UE}. 
{By exploiting the definition of payoff function} in~\eqref{payoff_function}, we can prove that $\mathcal{G}(\alpha)$ is a potential game by showing that it has an exact potential function as 
explained in the following result:
\begin{cor}
		\label{prop_2}
	There is an exact potential function for the parameterized game $\mathcal{G}(\alpha)$ and it is given by:
	\begin{align}
		\label{potential_function}
		u(\alpha, \mat{\rho}) &= \frac{1}{2}\sum_{k \in \mathcal{K}}\Delta \mu_k(\alpha, \mat{\rho}).
	\end{align}
	\begin{proof}
	This can be demonstrated with a relatively simple sequence of steps. 
	Let $\rho_{\tilde{k}}$, $\rho'_{\tilde{k}} \in \mathcal{P}_{\tilde{k}}$ be two different and arbitrary data power values for a generic \ac{UE} $\tilde{k}$. 
	Suppose that  \ac{UE} $\tilde{k}$ changes its data power from $\rho_{\tilde{k}}$ to $\rho'_{\tilde{k}}$, then the change of its payoff function is: $\Delta \mu_{\tilde{k}} = \mu_{\tilde{k}}(\alpha, \rho'_{\tilde{k}}, \mat{\rho}_{(-\tilde{k})})   - \mu_{\tilde{k}}(\alpha, \rho_{\tilde{k}}, \mat{\rho}_{(-\tilde{k})})$. 
	Moreover, the change of $u(\alpha, \mat{\rho})$ is:  $\Delta u = 	u(\alpha, \rho'_{\tilde{k}}, \mat{\rho}_{(-\tilde{k})}) - u(\alpha, \rho_{\tilde{k}}, \mat{\rho}_{(-\tilde{k})})$. By developing the expressions $\Delta \mu_{\tilde{k}} $ and $\Delta u$, it is possible to show that: $\Delta \mu_{\tilde{k}} = \Delta u$, and thus $u(\alpha, \mat{\rho})$ is an exact potential function for the parameterized game $\mathcal{G}(\alpha)$.
	\end{proof}
\end{cor}
\subsection{Proposed Iterative Algorithm}

In order to develop a potential game-based approach to address the problem of power control for the uplink of cell-free massive MIMO systems, we propose a procedure where the power allocation is updated every iteration for each \ac{UE} until reaching convergence. 
However, before specifically discussing this procedure, in order to achieve a practical implementation we define the following vector $\mat{\xi}\in \mathbb{R}^{K}$: $	\mat{\xi} = \left[\xi_{1}, \cdots, \xi_{k}, \cdots, \xi_{K}\right]^{\mathrm{T}}$,
where $\xi_{k} = \rho_k\Big(\sum\limits_{ l\in \mathcal{M}_k}\beta_{k,l}\Big)^\alpha, \, k \in \mathcal{K}$. Note that using the definition of $\mat{\xi}$, we can express the payoff function for each \ac{UE} in terms of $\mat{\xi}$ only, as follows: 
	\begin{align}
		\label{mu_xi}
		\mu_k(\mat{\xi}) = \frac{\sum\limits_{\forall i \neq k}\xi_{i}}{\xi_{k}} \,\, + \,\,\xi_{k}\sum_{\forall i \neq k}\frac{1}{\xi_{i}}.
	\end{align}

The complete procedure for the proposed power allocation strategy (PAS) based on game theory is described in Algorithm~\ref{ALG:GAME_ALGORITHM}. 
First, in line $1$, we can define $\mat{\rho}^{(0)}$ using any naive power allocation strategy, e.g., $\rho_k^{(0)}=P_{\text{max}}/n, \, k \in \mathcal{K}$, where $n \in \mathbb{R}^{*}_{+}$. 
Moreover, the input parameter $\alpha$ must also be initialized, which is an important variable for the game as will be discussed in Session~\ref{Simulation_Results_and_Discussion}.
\begin{algorithm}[h]
	\footnotesize
	\caption{\small Game-based power allocation strategy (Game-PAS)}\label{ALG:GAME_ALGORITHM}
	\begin{algorithmic}[1]
		\STATE 	\textbf{Input:} Initialize $\mat{\rho}^{(0)}$, $\alpha \in \mathbb{R}$, $l\leftarrow 0$, and $\epsilon>0$;
		\STATE 	\textbf{Output:} Data power vector $\mat{\rho}$;
		\LOOP
		\STATE $\text{UEs report the current data power to Master AP}$;
		\STATE {Master AP measures and broadcasts $\mat{\xi}^{(l)}$;}
		\STATE Find the payoff function using $\mat{\xi}^{(l)}$ and~\eqref{mu_xi};
		\STATE {Find $\rho^{\star}_{k}$ making $\rho^{\star}_{k} \leftarrow \min\{\rho^{*}_{k}, P_{\text{max}}\}$, where $\rho^{*}_{k}$ is defined in~\eqref{EQ:best_power}};
		\FOR {$k \in \mathcal{K}$}
		\IF{$\mu_{k}\big(\rho^{(l)}_{k},\mat{\rho}^{(l)}_{-(k)}\big) - \mu_{k}\big(\rho^{\star}_{k},\mat{\rho}^{(l)}_{-(k)}\big) > \epsilon$}
		\STATE $\rho^{(l+1)}_{k} \leftarrow \rho^{\star}_{k}$;
		\ELSE		
		\STATE $\rho^{(l+1)}_{k} \leftarrow \rho^{(l)}_{k}$;
		\ENDIF
		\ENDFOR
		$\,\,l \leftarrow l + 1$;
		\ENDLOOP \textbf{ when }$\rho^{(l)}_{k}= \rho^{(l-1)}_{k}, \,\, k \in \mathcal{K}$;	
	\end{algorithmic}
\end{algorithm}

In the \textit{outer loop} (lines $3$-$15$) the following procedure is repeated: after information exchange between the \acp{UE} and the Master \ac{AP}, each \ac{UE}  will independently choose a \textit{best response} to the actions of the other \acp{UE} {according to line $7$}. 
Then,  in case of an improvement of the payoff function,  each \ac{UE} updates its data power, otherwise the current data power value is kept. Analogously to the best response approach in line $7$, the power update also occurs individually, i.e., in the \textit{inner loop} (lines $8$-$14$) the power update for each \ac{UE} is performed in a distributed way and thus it does not depend on $K$. 
Finally, Algorithm~\ref{ALG:GAME_ALGORITHM} ends when no \ac{UE} can improve its payoff by unilateral deviation (cf. Definition~\ref{definition_1}).
\subsection{Signaling and Convergence Analysis}
\label{SUB_SEC:signaling}

During the execution of the proposed algorithm, an over-the-air signaling scheme must be considered for the two-way communication between the \acp{UE} and the Master \ac{AP}, as performed in the lines $4$ and $5$ of Algorithm~\ref{ALG:GAME_ALGORITHM}. 
Initially, each \ac{UE} reports to the Master \ac{AP} its current data power value. 
Next, the Master \ac{AP}  measures and then broadcasts $\mat{\xi}$. 
{This is how each user receives the power allocation from other users.}

Regarding the convergence of Algorithm~\ref{ALG:GAME_ALGORITHM}, all finite potential games have the finite improvement path property~\cite{mackenzie2006game}. Consequently, if every improvement path is finite and the best response approach provides an improvement of at least $\epsilon$, then it must necessarily converge~\cite[See Chapter 5, Theorem 19]{mackenzie2006game}. 
\section{Simulation Results and Discussion}
\label{Simulation_Results_and_Discussion}
In this section, we conduct simulations relying on the setup introduced in~\cite{Emil_Luca_2020}. 
In summary, the \acp{AP} and \acp{UE} are independently and uniformly distributed in a $ 2\text{ km}\times 2\text{ km}$ square. 
We apply the wrap-around technique to approximate an infinitely large network. 
Moreover, we assume that $\tau_{\text{c}}=200$, $\tau_{\text{p}}=10$, $\tau_{\text{u}}=190$, $P_{\text{max}} = 100\text{ mW}$ and a $20\text{ MHz}$ bandwidth. 
\subsection{Convergence Behavior}

\begin{figure}[h]
	\centering
	\includegraphics[width=.74\columnwidth]{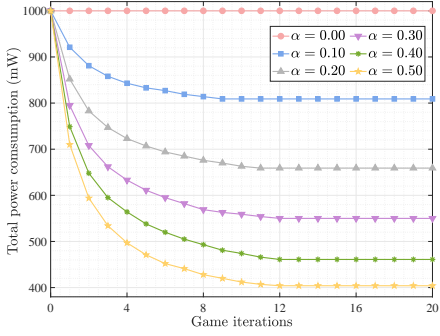}
	\caption{Impact of $\alpha$ on the total power consumption versus iterations of the game assuming $L=100$, $N=4$ and $K=10$.}
	\label{FIG:curve_convergence}
\end{figure}

Fig.~\ref{FIG:curve_convergence} shows the evolution of total power consumption of the \acp{UE} for different values of $\alpha$ versus the iterations of the game. 
For this result, we assume that initially all \acp{UE} transmit at their maximum power, i.e., $\rho_k^{(0)}=P_{\text{max}}, \, k \in \mathcal{K}$. 
Thus, the total power usage at the beginning of the game is $K\cdot P_{\text{max}} = 1000\text{ mW}$. 
In particular, when $\alpha = 0$ note that no \ac{UE} has an incentive to change the initial power allocation strategy and in that case the convergence of Algorithm~\ref{ALG:GAME_ALGORITHM} is immediate. 

On the other hand, in a more general case when $\alpha\neq 0$, the convergence behavior is different. 
As $\alpha$ increases, the strategy followed by the \acp{UE} converges to situations where the power expenditure is decreasing, i.e., the \acp{UE} are turned to a low power mode. 
As a result, this allows to improve the use of the energy stored in the battery and it shows that, interestingly, $\alpha$ has a direct impact on energy-saving. {Obviously, as the values of $\alpha$ vary, we also obtain different data rates for the UEs} and, therefore, non-trivial solutions especially in terms of \ac{EE} {defined as bit/J} can be obtained.

\subsection{Performance Comparison}

In this section, the performance evaluation of the proposed power control is evaluated based on two aspects. 
First, we show the performance of Algorithm~\ref{ALG:GAME_ALGORITHM} in three different scenarios, namely, cell-free (discussed in Section~\ref{Sec:Network_Model}), small cell and massive MIMO systems. This is interesting as it shows the good adaptability of game theory-based approaches to various frameworks. 
In each scenario, we also consider a baseline scheme, in which 
each \ac{UE} transmits at full power, i.e., we use the greedy power allocation strategy (Greedy-PAS)
as benchmarking. 
It has been shown in the literature that this power allocation strategy can provide good \ac{SE} and fairness~\cite{Emil_Luca_2020}. 
Furthermore, we consider three different metrics: the total system \ac{SE}, the minimum \ac{SE}, and the total system \ac{EE}, which is the sum of the EEs of each \ac{UE}, defined as the ratio between the \ac{SE} and the corresponding consumed power. Finally, our power control (Game-PAS) is performed for different $\alpha$ within the range $[0,2]$ and the best performance for each metric is depicted.

\begin{figure}[t]
	\centering
	\includegraphics[width=.74\columnwidth]{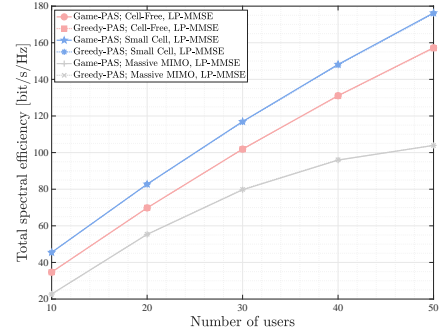}
	\caption{Total spectral efficiency versus number of users assuming $L=100, N=4$ for the cell-free/small cell setups and $L=4, N=100$ for the massive MIMO case.}
	\label{FIG:total_SE_vs_users}
\end{figure}

Fig.~\ref{FIG:total_SE_vs_users} plots the total spectral efficiency versus the number of users. 
Specifically in this metric, the performance obtained by the proposed and baseline solutions are the same in all simulated setups. 
Basically, it means that from the point of view of total SE, and due to its simpler implementation, the Greedy-PAS solution has a better trade-off between performance and computational cost and is, therefore, the best option.
\begin{figure}[h]
	\centering
	\includegraphics[width=.74\columnwidth]{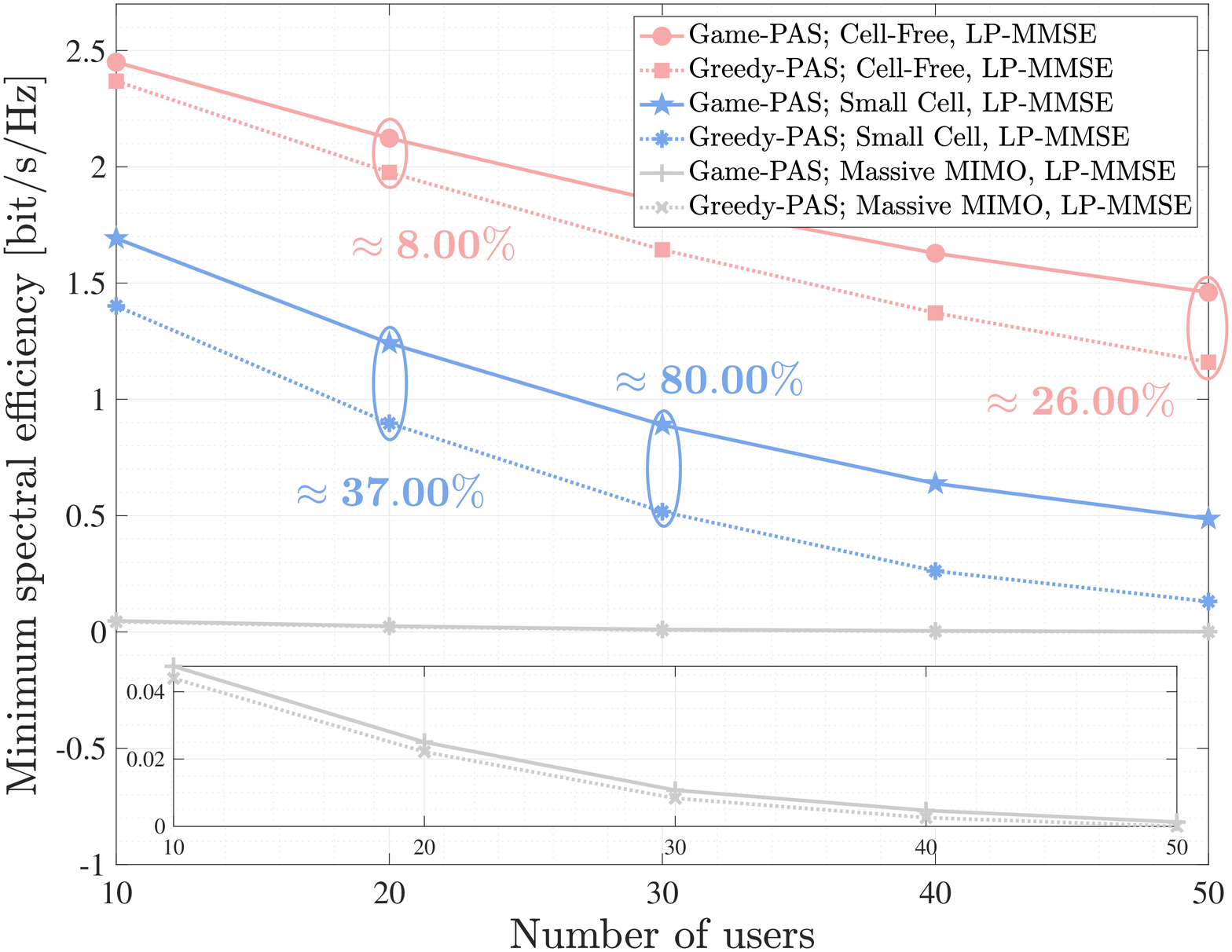}
	\caption{Minimum spectral efficiency versus number of users assuming $L=100, N=4$ for the cell-free/small cell setups and $L=4, N=100$ for the massive MIMO case.}
	\label{FIG:min_SE_vs_users}
\end{figure}

On the other hand, for the cell-free and small cell setups, significant performance gains in terms of minimum \ac{SE} can be achieved using the proposed power control, as shown in Fig.~\ref{FIG:min_SE_vs_users}.
Moreover, we highlight that the gains tend to increase as the number of \acp{UE} increases. 
For the cell-free case, for example, we have average percentage gains around $8\%$ and $26\%$ when $K=20$ and $K=50$, respectively. 
Note that even more expressive gains of the proposed solution are obtained for the small cell case. 
In general, under interference-limited environments, as the number of \acp{UE} in the system increases, the power control problem becomes more relevant. 
However, trivial power allocation strategies usually neglect the impact of increasing \acp{UE} and, consequently, are ineffective in mitigating network interference by means of power control. 

\begin{figure}[t]
	\centering
	\includegraphics[width=.74\columnwidth]{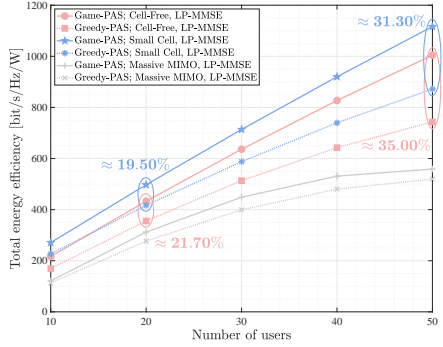}
	\caption{Total energy efficiency versus number of users assuming $L=100, N=4$ for the cell-free/small cell setups and $L=4, N=100$ for the massive MIMO case.}
	\label{FIG:total_EE_vs_users}
\end{figure}

Finally, we plot the \ac{EE} in Fig.~\ref{FIG:total_EE_vs_users}. 
First, we highlight that the impact of $\alpha$ on the power usage shown in Fig.~\ref{FIG:curve_convergence} has a direct effect in achieving enhanced \acp{EE}. 
Further, note that similarly to the minimum \ac{SE} metric, the total \ac{EE} performance gains also increase as $K$ increases. 
This is particularly interesting as increasing the number of \acp{UE} in the network can rapidly lead to a growing concern with excessive energy demand, especially for the Greedy-PAS solution. 
At this point, energy efficient solutions are important and a more robust power allocation strategy such as the Game-PAS is critical to reduce the energy cost per transmitted bit and to improve the greenness of wireless systems.

\subsection{Trade-off between EE and SE}
\begin{figure}[h]
	\centering
	\includegraphics[width=.76\columnwidth]{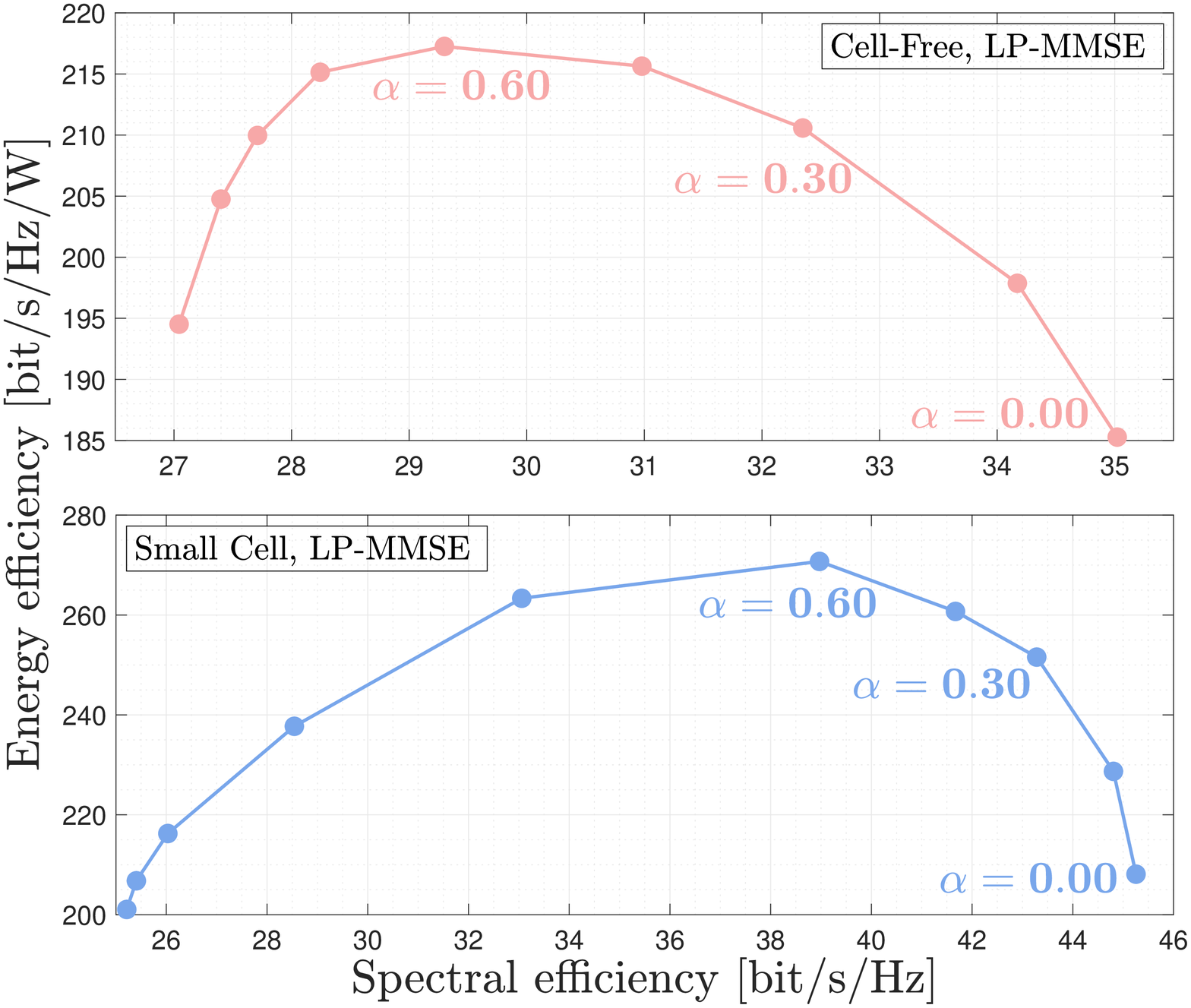}
	\caption{Trade-off curve between EE and SE for the cell-free and small cell setups assuming $L=100, N=4$ and $K=10$.}
	\label{FIG:trade_off_EE_SE}
\end{figure}
It is well-known that \ac{EE} and \ac{SE} are conflicting objectives and there exists an inherent trade-off between them. 
Thus, in the context of the proposed solution, it is interesting to show the impact of parameter $\alpha$ on the \ac{EE}-\ac{SE} trade-off. 
Fig.~\ref{FIG:trade_off_EE_SE} presents the achieved \ac{SE} and \ac{EE} with different values of $\alpha$ for the cell-free and small cell setups. 
From the point of view of maximizing the \ac{SE}, when $\alpha = 0.00$, as discussed in Fig.~\ref{FIG:curve_convergence}, the Game-PAS solution is equivalent to the Greedy-PAS solution and, in this case, the systems achieve high \ac{SE}. 
However, as $\alpha$ increases, the \ac{EE} is gradually improved until reaching a maximum value when $\alpha = 0.60$. 
Also, for other values of $\alpha$, different solutions for \ac{EE} and \ac{SE} can be obtained. 
Therefore, Fig.~\ref{FIG:trade_off_EE_SE} demonstrates that the proposed solution is efficient in achieving a flexible trade-off between \ac{EE} and \ac{SE}. 
For example, for small cells, when $\alpha = 0.30$, the \ac{EE} metric has a gain around $20\%$ with a small cost in terms of \ac{SE}.

\section{Final Comments}
In this paper, we proposed a distributed game-theoretic method for power control in the uplink of cell-free systems. 
Simulations indicate that the proposed solution achieves significant performance gains in terms of minimum \ac{SE} floor and power consumption with an improved \ac{EE}. 
Moreover, by varying the $\alpha$, we showed that it is possible to achieve different solutions for \ac{EE} and \ac{SE}. 
Hence, the proposed solution simplifies the process of joint optimization of these metrics and allows to obtain useful trade-offs between \ac{EE} and \ac{SE}.

\section{Acknowledgment}

This work was supported in part by Ericsson Research,
Technical Cooperation Contract UFC.48, in part by CNPq,
in part by FUNCAP, in part by CAPES/PRINT
Grant 88887.311965/2018-00 and by CAPES - Finance
Code 001.

\ifCLASSOPTIONcaptionsoff
  \newpage
\fi

\bibliographystyle{IEEEtran}
\bibliography{IEEEabrv,biblio}

\end{document}